\documentclass[aps, prl, twocolumn, superscriptaddress, amsmath, showpacs, tightenlines, footinbib, longbibliography]{revtex4-1}
\usepackage[colorlinks, breaklinks, citecolor=blue, linkcolor=blue,  urlcolor={blue}]{hyperref}
\usepackage{breakurl}
\usepackage{amsfonts}
\usepackage{amsmath}
\usepackage{amssymb}
\usepackage{mathrsfs}
\usepackage{epsfig}
\usepackage{graphicx}
\usepackage{bm}

\begin{document}

\title{Loss-Induced Quantum Revival}
\author{Yunlan Zuo}\thanks{Co-first authors with equal contribution}
\affiliation{Key Laboratory of Low-Dimensional Quantum Structures and Quantum Control of Ministry of Education, Department of Physics and Synergetic Innovation Center for Quantum Effects and Applications, Hunan Normal University, Changsha 410081, China}
\author{Ran Huang}\thanks{Co-first authors with equal contribution}
\affiliation{Key Laboratory of Low-Dimensional Quantum Structures and Quantum Control of Ministry of Education, Department of Physics and Synergetic Innovation Center for Quantum  Effects and Applications, Hunan Normal University, Changsha 410081, China}
\affiliation{Theoretical Quantum Physics Laboratory, RIKEN Cluster for Pioneering Research, Wako-shi, Saitama 351-0198, Japan}
\author{Le-Man Kuang}
\email{lmkuang@hunnu.edu.cn}
\affiliation{Key Laboratory of Low-Dimensional Quantum Structures and Quantum Control of Ministry of Education, Department of Physics and Synergetic Innovation Center for Quantum Effects and Applications, Hunan Normal University, Changsha 410081, China}
\author{Xun-Wei Xu}
\email{davidxu0816@163.com}
\affiliation{Key Laboratory of Low-Dimensional Quantum Structures and Quantum Control of Ministry of Education, Department of Physics and Synergetic Innovation Center for Quantum Effects and Applications, Hunan Normal University, Changsha 410081, China}
\author{Hui Jing}
\email{jinghui73@foxmail.com}
\affiliation{Key Laboratory of Low-Dimensional Quantum Structures and Quantum Control of Ministry of Education, Department of Physics and Synergetic Innovation Center for Quantum  Effects and Applications, Hunan Normal University, Changsha 410081, China}
\date{\today}

\begin{abstract}
Conventional wisdom holds that quantum effects are fragile and can be destroyed by loss. Here, contrary to general belief, we show how to realize quantum revival of optical correlations at the single-photon level with the help of loss. We find that, accompanying loss-induced transparency of light in a nonlinear optical-molecule system, quantum suppression and revival of photonic correlations can be achieved. Specifically, below critical values, adding loss into the system leads to suppressions of both optical intensity and its non-classical correlations; however, by further increasing the loss beyond the critical values, quantum revival of photon blockade (PB) can emerge, resulting in loss-induced switch between single-PB and two-PB or super-Poissonian light. Our work provides a strategy to reverse the effect of loss in fully quantum regime, opening up a counterintuitive route to explore and utilize loss-tuned single-photon devices for quantum technology.
\end{abstract}

\maketitle

Loss is ubiquitous in nature, which is usually regarded as harmful and undesirable in making and operating quantum devices. Very recently, loss has been found to play an unconventional role in non-Hermitian physics~\cite{bender2007Making,rotter2009NonHermitian,el-ganainy2019Dawn,ozdemir2019Parity}, such as loss-induced transparency of light~\cite{Guo2009Observation,Zhang2018Loss}, loss-induced lasing revival~\cite{peng2014Lossinduced}, and loss-induced nonreciprocity~\cite{Huang2021Loss,Dong2020Loss}. These pioneering works, however, have mainly focused on the classical regime, i.e., studying loss-tuned optical intensity, instead of quantum correlation of light. Understanding the role of loss in engineering purely quantum effects not only facilitates the development of open quantum theories, but also provides a practical way to fabricate loss-controlled quantum devices inaccessible by conventional ways and allows exploring their applications in quantum technology.

In this work, as a step towards this goal, we show how to achieve quantum revival of a purely quantum effect, i.e., photon blockade (PB), with the help of loss. PB, showing photons behave as effectively impenetrable particles, has been demonstrated in diverse systems ranging from cavity QED~\cite{birnbaum2005Photon,faraon2008Coherent,muller2015Coherent,hamsen2017TwoPhoton,Snijders2018Observation} to superconducting circuits~\cite{Vaneph2018Observation,lang2011Observation,hoffman2011Dispersive} and cavity free devices~\cite{Peyronel2012Quantum}. PB provides a unique way not only to make important quantum devices~\cite{leonski1994possibility,imamoglu1997Strongly,rabl2011Photon,liao2013Photon,liao2013Correlated,lu2015Squeezed,wang2015Tunable,zhu2018Controllable,zou2019Enhancement,zhai2019Mechanical,shamailov2010Multiphoton,miranowicz2013Twophoton,bin2018Twophoton,ghosh2019Dynamical,Roberts2020Driven,You2020Reconfigurable}, such as single-photon turnstiles~\cite{Dayan2008A}, single-photon routers~\cite{Shomroni2014All}, or quantum circulators~\cite{Scheucher2016Quantum}, but also to explore the fundamental issues of quantum many-body physics~\cite{Jin2013Photon,Greentree2006Quantum,Angelakis2007Photon,Noh2016Quantum,zeytinoglu2018Interactioninduced,pietikainen2019Photon,Kyriienko2020Nonlinear,Iversen2021Strongly}. To date, the main approaches for realizing PB can be classified into two groups: strong-nonlinearity-induced anharmonicity in energy spectrum of the system~\cite{birnbaum2005Photon,faraon2008Coherent,muller2015Coherent,hamsen2017TwoPhoton,lang2011Observation,hoffman2011Dispersive,Peyronel2012Quantum,ridolfo2012Photon,majumdar2013Singlephoton,liu2014Blockade,huang2018Nonreciprocal}, and destructive interference between different modes~\cite{Liew2010Single,Bamba2011Origin,Majumdar2012Loss,Flayac2017Unconventional,Snijders2018Observation,Vaneph2018Observation,zou2020Enhancement,li2019Nonreciprocal}.
Generically, in both cases, the optical loss should be smaller than the strength of nonlinearity or coupling of different modes, since it is regarded as limiting the efficiency or functionalities of PB devices.

Here we show that, accompanying the classical revival of optical intensities, quantum correlations of light can also be revived by adding loss in an optical compound system. We note that in the pioneering experiments on loss-induced transparency~\cite{Guo2009Observation,peng2014Lossinduced}, classical suppression and revival of optical transmission are attributed to the emergence of an exceptional point (EP), featuring the coalescence of both the complex eigenvalues and their corresponding eigenstates~\cite{heiss2004Exceptional,Miri2019exceptional}. In contrast, we find that in our system, quantum suppression and revival of optical correlations precisely correspond to the conditions of two-photon resonance and excitation-spectrum mode coalescence. More interestingly, we also find that different types of quantum correlations can emerge in the revived light, by merely increasing the loss (via placing an external nanotip near the optical resonator), resulting in loss-tuned quantum switches between single-photon and two-photon blockades. Our work extends loss-induced effects into the purely quantum regime, opening up a promising way to study various quantum effects with lossy synthetic materials~\cite{Feng2012Loss,Dong2020Loss} or topological structures~\cite{Fesenko2019Lossless,Qiao2021Higher}, as well as to build loss-tuned single-photon devices for quantum engineering~\cite{harris1998photon, chang2007single,kubanek2008two} and quantum metrology~\cite{fattal2004entanglement,buluta2009quantum,georgescu2014quantum}.

\begin{figure}[tb]
\centering
\includegraphics[width=0.48 \textwidth]{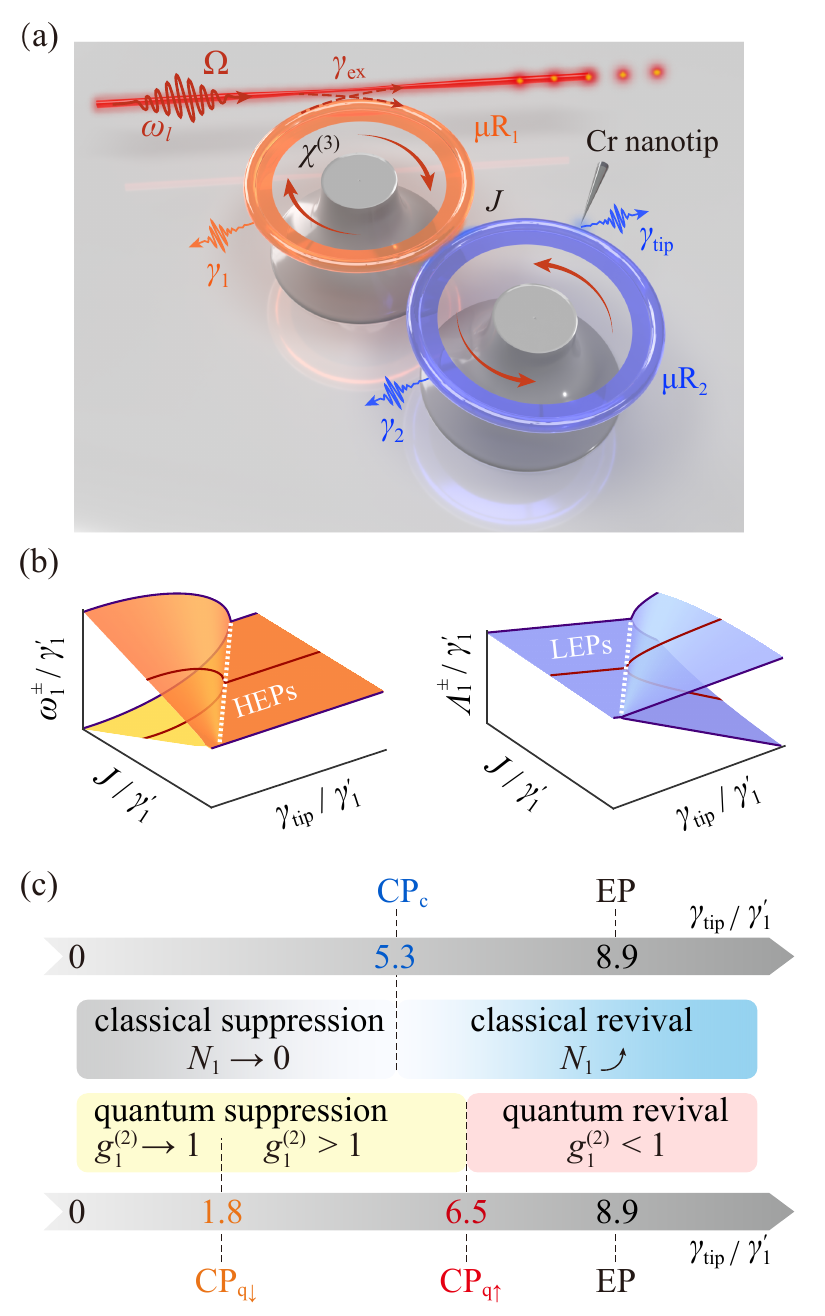}
\caption{Loss-induced suppression and revival in an optical compound system. (a) A whispering-gallery-mode resonator $\mu\mathrm{R1}$ with Kerr-type nonlinearity $\chi$ coupled to a linear optical cavity $\mu\mathrm{R2}$ with additional loss $\gamma_{\mathrm{tip}}$ induced by a Cr-coated nanotip. (b) The locations of Hamiltonian exceptional points (HEPs, black dashed line) agree well with those of Liouvillian exceptional points (LEPs) obtained through the fully quantum simulations~\cite{SM}. Here, we focus on the case of $J/\gamma_1^{'}=2$ (red solid curves). (c) The EP at $\gamma_{\mathrm{tip}}/\gamma_1^{'}=8.9$ leads to classical and quantum critical points, $\mathrm{CP}_\mathrm{c,q}$, in mean photon number $N_1$ and quantum correlation $g^{(2)}(0)$, respectively. Here, $\mathrm{CP}_{\mathrm{q}\downarrow}$ and $\mathrm{CP}_{\mathrm{q}\uparrow}$ are related to the quantum suppressive and revived processes, respectively. For the experimentally accessible parameter values, see the main text.} \label{fig:FP1}
\end{figure}

\begin{figure*}[tb]
\centering
\includegraphics[width=0.96 \textwidth]{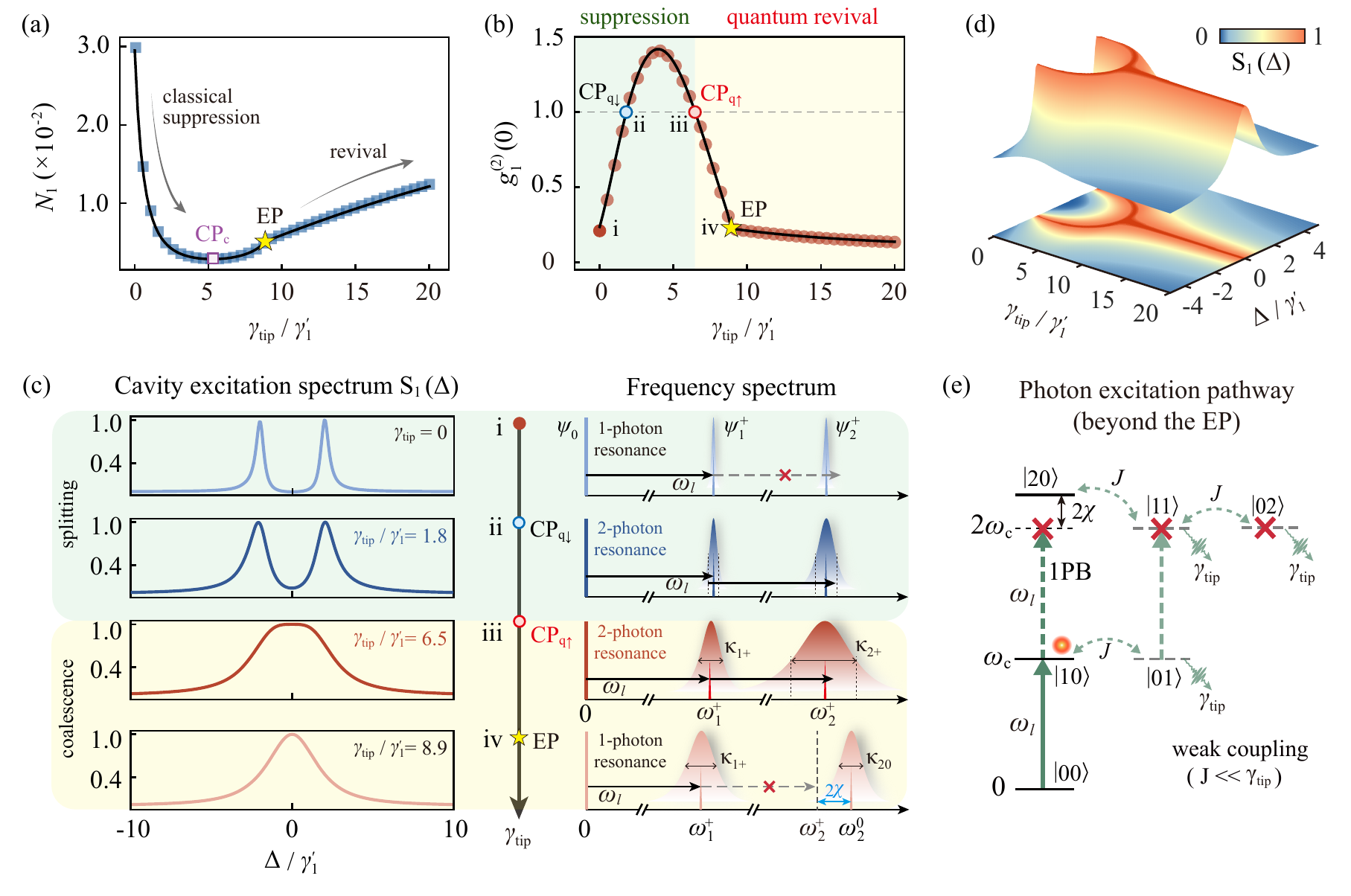}
\caption{(a) The intracavity photon number $N_{1}$ and (b) the second-order correlation function $g^{(2)}_1(0)$ versus $\gamma_{\mathrm{tip}}$. The markers (squares, circles) and black lines are analytical and numerical solutions, respectively. (c) The cavity excitation spectrum $S_{1}(\Delta)$ and the eigenfrequency spectra $\omega_{1,2}$ with linewidths $\kappa_{1,2}$ show the origin of the quantum suppression and revival. (d) The revived 1PB can be understood from the mode coalescence in $S_{1}(\Delta)$, and (e) the photon excitation pathway beyond the EP, where $\omega_1^\pm$ coalesce at $\omega_c$, and the dashed arrows are the forbidden excitations. The parameters are the same as those in Fig.~\ref{fig:FP1}.} \label{fig:FP2}
\end{figure*}

We consider a single-PB device consisting of an optical Kerr resonator ($\mu\mathrm{R1}$) directly coupled to a linear optical resonator ($\mu\mathrm{R2}$) through evanescent fields, with the coupling strength $J$, as shown in Fig.~\ref{fig:FP1}(a). The system without driving is described by the Hamiltonian $(\hbar=1)$:
\begin{align}\label{eq:H}
\hat{H}_{\mathrm{i}} =\sum_{j=1,2}\omega_{c}\hat{a}_{j}^{\dagger}\hat{a}_{j}+\chi\hat{a}_{1}^{\dagger}\hat{a}_{1}^{\dagger}\hat{a}_{1}\hat{a}_{1}+J(\hat{a}_{1}^{\dagger}\hat{a}_{2}+\hat{a}_{2}^{\dagger}\hat{a}_{1}),
\end{align}
where $\hat{a}_{j=1,2}$ are the intracavity modes with resonance frequency $\omega_c$, and $\chi=3\hbar\omega_c^2\chi^{(3)}/(4\varepsilon_0\varepsilon_r^2V_\mathrm{eff})$ is the Kerr parameter with vacuum (relative) permittivity $\varepsilon_0$ ($\varepsilon_r$), nonlinear susceptibility $\chi^{(3)}$, and mode volume $V_\mathrm{eff}$. In addition to highly nonlinear materials~\cite{hales2018third,Heuck2020Controlled,Alam2016Large,zielinska2017self,Choi2017Self}, Kerr-type nonlinearity can also be achieved in cavity or circuit QED systems~\cite{birnbaum2005Photon,Kirchmair2013Observation,gu2017microwave}, cavity free systems~\cite{Xia2018Cavity}, as well as optomechanical~\cite{Gong2009Effective,rabl2011Photon,lu2013quantum} or magnon devices~\cite{wang2018bistability,Zhang2021Exceptional}.

The intrinsic losses of the two resonators are $\gamma_{j=1,2}$. The total loss of $\mu\mathrm{R1}$ is given by $\gamma_{1}^{'}=\gamma_{1}+\gamma_\mathrm{ex}$, where $\gamma_\mathrm{ex}$ is the loss induced by the coupling between the resonator and the fiber taper. An additional loss $\gamma_{\mathrm{tip}}$ is introduced on $\mu\mathrm{R2}$ by a chromium (Cr) coated silica-nanofiber tip, featuring strong absorption in the $1550\ \mathrm{nm}$ band~\cite{peng2014Lossinduced}. The strength of $\gamma_{\mathrm{tip}}$ can be increased by enlarging the volume of the nanotip within the linear cavity mode field, leading to a linewidth broadening without observable change in resonance frequency~\cite{peng2014Lossinduced}. Thus, the total loss of $\mu\mathrm{R2}$ is given by $\gamma_{2}^{'}=\gamma_{2}+\gamma_{\mathrm{tip}}$.

We study the eigenenergy spectrum of this system by considering the effects of loss. The eigenstates $|\psi_{1,2}^{\pm,0}\rangle$ are the superposition states of the Fock state $|m,n\rangle$ with $m$ photons in $\mu\mathrm{R1}$ and $n$ photons in $\mu\mathrm{R2}$~\cite{SM}. The complex eigenvalues of this non-Hermitian system in the one-photon excitation subspace are found as
\begin{align}\label{eq:N1}
\lambda_{1}^{\pm}=-i\Gamma+\omega_{c}\pm\sqrt{J^{2}-\beta^{2}},
\end{align}
whose real and imaginary parts are respectively indicate the eigenfrequencies $\omega_{1}^{\pm}$ and the linewidths $\kappa_{1}^{\pm}$. Here, $\Gamma=(\gamma_{1}^{'}+\gamma_{2}^{'})/{4}$ and $\beta=(\gamma_{2}^{'}-\gamma_{1}^{'})/{4}$ quantify the total loss and the loss contrast of the system, respectively.

The Hamiltonian EPs (HEPs) are defined as the spectral degeneracies of the non-Hermitian Hamiltonian~\cite{heiss2004Exceptional,Miri2019exceptional}, which emerge for $\lambda_{1}^{+}=\lambda_{1}^{-}$, i.e.,
\begin{align}\label{eq:HEPs}
\gamma_{\mathrm{tip}}^{\mathrm{EP}}=4J+\gamma_{1}^{'}-\gamma_{2}.
\end{align}
For a full quantum picture, we study Liouvillian EPs (LEPs) including the effect of quantum jumps~\cite{Minganti2019quantum,SM}. As shown in Fig.~\ref{fig:FP1}(b), the LEPs and HEPs occur at the same positions indicating a good agreement between the semiclassical and fully quantum approaches~\cite{Minganti2019quantum}.

As what one would expect in conventional systems, additional loss $\gamma_{\mathrm{tip}}$ decreases the mean-photon number $N_1$ to zero in $\mu\mathrm{R1}$. However, $N_1$ recovers with more loss in the vicinity of the classical critical point ($\mathrm{CP}_\mathrm{c}$), i.e., the $\gamma_{\mathrm{tip}}$ with the minimum of $N_1$ [Fig.~\ref{fig:FP1}(c)]. The quantum statistics of this light can be recognized from the second-order correlation function $g^{(2)}_1(0)$. The condition $g^{(2)}_1(0)<1$ [$g^{(2)}_1(0)>1$] characterizes sub-Poissonian (super-Poissonian) statistics or photon antibunching (bunching), and $g^{(2)}_1(0)\to0$ indicates a full single-PB. Adding loss annihilates the single-PB, and converts the light from antibunching into bunching. We refer to the $\gamma_{\mathrm{tip}}$ for $g^{(2)}_1(0)=1$ as quantum critical points ($\mathrm{CP}_{\mathrm{q}\downarrow,\uparrow}$). Remarkably, in the vicinity of $\mathrm{CP}_{\mathrm{q}\uparrow}$, the sub-Poissonian light recovers despite the increasing loss, with the revival of single-PB at an EP. More intriguingly, when $N_1$ recovers after $\mathrm{CP}_\mathrm{c}$, the quantum statistics of the light can be tuned between bunching and antibunching by increasing loss below or beyond $\mathrm{CP}_{\mathrm{q}\uparrow}$, respectively. This loss-induced quantum revival is fundamentally different from the classical revival of transmission rates~\cite{Guo2009Observation,Zhang2018Loss,peng2014Lossinduced}.

To study this loss-induced quantum revival, we consider the Hamiltonian $\hat{H}_{\mathrm{i}}$ in a frame rotating with the driving frequency $\omega_{l}$: $\hat{H}_{\mathrm{r}} =\sum_{j=1,2}\Delta\hat{a}_{j}^{\dagger}\hat{a}_{j}+\chi\hat{a}_{1}^{\dagger}\hat{a}_{1}^{\dagger}\hat{a}_{1}\hat{a}_{1}+J(\hat{a}_{1}^{\dagger}\hat{a}_{2}+\hat{a}_{2}^{\dagger}\hat{a}_{1})+\Omega(\hat{a}_{1}^{\dagger}+\hat{a}_{1})$, where $\Delta=\omega_{c}-\omega_{l}$ is the optical detuning, $\Omega=[{\gamma_{\mathrm{ex}}P_{\mathrm{in}}/(\hbar\omega_{l})}]^{1/2}$ is the driving amplitude with power $P_{\mathrm{in}}$ on $\mu\mathrm{R1}$. The optical decay can be included in the effective Hamiltonian $\hat{H}_{\mathrm{eff}}=\hat{H}_{\mathrm{r}}-i{\sum}_{j=1,2}(\gamma_{j}^{'}/2)\hat{a}_{j}^{\dagger}\hat{a}_{j}$~\cite{plenio1998quantum}. The probabilities of finding $m$ photons in $\mu\mathrm{R1}$ and $n$ photons in $\mu\mathrm{R2}$ are given by $P_{mn}=|C_{mn}|^{2}$ with probability amplitudes $C_{mn}$, which can be solved through Schr\"odinger equation~\cite{SM}. For weak driving $(\Omega\ll\gamma_{1}^{'})$, by truncating the Hilbert space to $N=m+n=3$, the mean-photon number in $\mu\mathrm{R1}$ is:
\begin{align}\label{eq:N1}
N_{1}=\langle \hat{a}_{1}^{\dagger}\hat{a}_{1}\rangle=\sum_{N=0}^3\sum_{m=0}^NmP_{mn},
\end{align}
and the equal-time second-order correlation function is
\begin{align}\label{eq:GTA}
g^{(2)}_1(0)=\frac{\langle\hat{a}_{1}^{\dagger2}\hat{a}_{1}^{2}\rangle}{\langle\hat{a}_{1}^{\dagger}\hat{a}_{1}\rangle^2}\simeq\frac{2P_{20}}{N_{1}^{2}}.
\end{align}

In order to confirm our analytical results, we numerically study the full quantum dynamics of the system. We introduce the density operator $\hat{\rho}(t)$ and then solve the master equation~\cite{Johansson2012QuTiP,Johansson2013QuTiP}
\begin{align}\label{eq:ME}
\dot{\hat{\rho}}=-i[\hat{H}_{\mathrm{r}},\hat{\rho}]+\underset{j=1,2}{\sum}\frac{\gamma_{j}^{'}}{2}(2\hat{a}_{j}\hat{\rho}\hat{a}_{j}^{\dagger}-\hat{a}_{j}^{\dagger}\hat{a}_{j}\hat{\rho}-\hat{\rho}\hat{a}_{j}^{\dagger}\hat{a}_{j}).
\end{align}
Then, $P_{mn}=\langle m,n|\rho_\text{ss}|m,n\rangle$ can obtained from the steady-state
solutions $\rho_\text{ss}$ of this master equation. The experimentally accessible parameters are chosen as~\cite{vahala2003optical,spillane2005ultrahigh,pavlov2017pulse,huet2016millisecond,hales2018third,Heuck2020Controlled,Alam2016Large,zielinska2017self,Choi2017Self,schuster2008nonlinear}: $V_\mathrm{eff}=100\ \mu\mathrm{m}^3$, $Q=2\times10^9$, $\chi^{(3)}/\varepsilon_r^2=2\times10^{-17}\ \mathrm{m}^2/\mathrm{V}^2$, $P_\mathrm{in}=4\ \mathrm{fW}$, $\lambda=1550\ \mathrm{nm}$. For the whispering-gallery-mode resonators, $V_\mathrm{eff}$ is typically $10^2$--$10^4\ \mu\mathrm{m}^3$~\cite{vahala2003optical,spillane2005ultrahigh}, and $Q$ has been increased up to $10^9$--$10^{12}$~\cite{pavlov2017pulse,huet2016millisecond}. The Kerr coefficient can be $\chi^{(3)}/\varepsilon_r^2=2\times10^{-17}\ \mathrm{m}^2/\mathrm{V}^2$ for the semiconductor materials with GaAs~\cite{hales2018third,Heuck2020Controlled}, and reach $\chi^{(3)}/\varepsilon_r^2=2.12\times10^{-17}\ \mathrm{m}^2/\mathrm{V}^2$ for the materials with indium tin oxide~\cite{Alam2016Large}. In addition, $\chi^{(3)}$ can be further enhanced to $2\times10^{-11}\ \mathrm{m}/\mathrm{V}^2$ by introducing other materials~\cite{zielinska2017self,Choi2017Self}.

An excellent agreement between our analytical results and the exact numerical results is seen in Fig.~\ref{fig:FP2}. Figure~\ref{fig:FP2}(a) shows the loss-induced classical suppression and revival of the intracavity photon number $N_{1}$. Below $\mathrm{CP}_\mathrm{c}$, $\gamma_{\mathrm{tip}}/\gamma_1^{'}=5.3$, $N_{1}$ is decreased to $0.003$ by increasing additional loss. When the loss exceeds $\mathrm{CP}_\mathrm{c}$, $N_{1}$ is revived due to the EP-induced mode coalescence; resulting in a predominant mode localized in $\mu\mathrm{R1}$. This classical counterintuitive effect has been used for realizing loss-induced revival of lasing~\cite{peng2014Lossinduced}.

\begin{figure}[t]
\centering
\includegraphics[width=0.48 \textwidth]{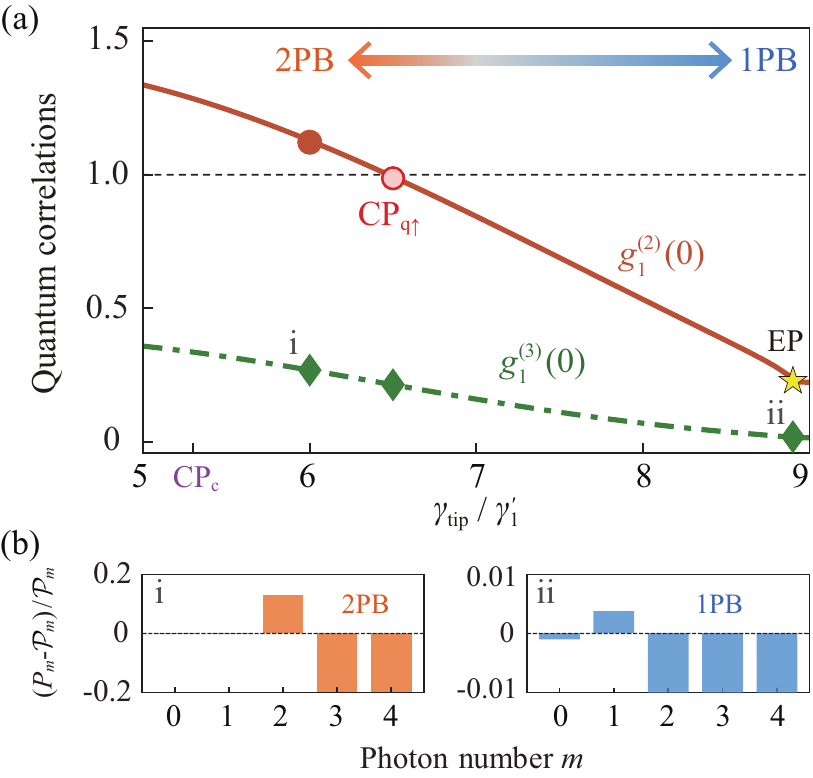}
\caption{Loss-induced quantum switching between two-photon blockade (2PB) and single-photon blockade (1PB). (a) Quantum correlations $g^{(2)}_1(0)$ (red solid curve) and $g^{(3)}_1(0)$ (green dashed curve) versus $\gamma_{\mathrm{tip}}$. (b) This loss-induced quantum switching can also be recognized from the deviations of the photon distribution $P_m$ to the standard Poisson distribution $\mathcal{P}_{m}$ with the same mean photon number $m$. The parameters are the same as those in Fig.~\ref{fig:FP2}. } \label{fig:FP3}
\end{figure}

More importantly, we find a loss-induced quantum revival of single-PB in Fig.~\ref{fig:FP2}(b). For $\gamma_{\mathrm{tip}}=0$, single-PB emerges with $g^{(2)}_1(0)\sim0.23$. Adding loss annihilates the single-PB, where the sub-Poissonian light is converted into coherent stream on $\mathrm{CP}_{\mathrm{q}\downarrow}$ ($\gamma_{\mathrm{tip}}/\gamma_1^{'}=1.8$), and turned into super-Poissonian light with $g^{(2)}_1(0)\sim1.42$ for more loss. Surprisingly, the sub-Poissonian light recovers by further increasing loss beyond $\mathrm{CP}_{\mathrm{q}\uparrow}$ ($\gamma_{\mathrm{tip}}/\gamma_1^{'}=6.5$), and single-PB is fully revived on the EP ($\gamma_{\mathrm{tip}}/\gamma_1^{'}=8.9$).

The loss-induced quantum suppression and revival require the interplay of mode coalescence and two-photon resonance [Fig.~\ref{fig:FP2}(c)]. The excitation spectrum $S_{1}(\Delta)=N_1/n_0$, with $n_{0}=\Omega^{2}/(\gamma_{1}^{'}+\gamma_{2}^{'})^{2}$, shows the mode splitting and coalescence [Figs.~\ref{fig:FP2}(c) and \ref{fig:FP2}(d)]. Below $\mathrm{CP}_{\mathrm{q}\uparrow}$, two spectrally separated modes are seen in Fig.~\ref{fig:FP2}(c-i,ii). The light with frequency $\omega_1^+$ is resonantly coupled to the transition $|\psi_{0} \rangle \rightarrow |\psi_{1}^{+} \rangle$, while $|\psi_{1}^{+} \rangle \rightarrow |\psi_{2}^{+} \rangle$ is detuned, resulting in a single-PB at $\gamma_{\mathrm{tip}}=0$. By further adding $\gamma_{\mathrm{tip}}$ to $\mathrm{CP}_{\mathrm{q}\downarrow}$, the light coincides with the two-photon resonance, leading to a suppression of single-PB.

The two-photon resonance remains for adding loss from $\mathrm{CP}_{\mathrm{q}\downarrow}$ to $\mathrm{CP}_{\mathrm{q}\uparrow}$ [Fig.~\ref{fig:FP2}(c-iii)]. However, increasing $\gamma_{\mathrm{tip}}$ to $\mathrm{CP}_{\mathrm{q}\uparrow}$ leads to an overlap of the mode resonances. Eventually, the modes coalesce at the EP [Fig.~\ref{fig:FP2}(c-iv)], indicating the coupled cavities entered the weak-coupling regime ($J\ll\gamma_{\mathrm{tip}}$)~\cite{peng2014Lossinduced}.

This mode coalescence can break the condition of two-photon resonance resulting in a quantum revival of single-PB. Specifically, two-photon eigenstates $|\psi_{2}^{\pm} \rangle$ are intensively localized on $|0,2\rangle$ and $|1,1\rangle$ by increasing loss beyond the EP~\cite{SM}. Although $|0,2\rangle$ or $|1,1\rangle$ coincides with the two-photon resonance energy $2\omega_c$, the two-photon resonance transitions from $|0,0\rangle$ to $|0,2\rangle$ and $|1,1\rangle$, i.e., $|\psi_{0} \rangle \rightarrow |\psi_{2}^{\pm} \rangle$, are forbidden due to the EP-induced mode coalescence and the effective weak coupling between the two cavities [Fig.~\ref{fig:FP2}(e)].

In addition, $|\psi_{2}^{0} \rangle$ and $|\psi_{1}^{+} \rangle$ are respectively governed by the states $|2,0\rangle$ and $|1,0\rangle$ when the system operates at or beyond the EP. As shown in Fig.~\ref{fig:FP2}(e), when the light resonantly coupled to $|0,0\rangle \rightarrow |1,0\rangle$, the transition from $|1,0\rangle$ to $|2,0\rangle$ is detuned by $2\chi$, indicating a single-PB is revived because of the anharmonic energy-level spacing induced by Kerr nonlinearity. We conclude that the interplay of excitation-spectrum mode coalescence and the two-photon resonance in nonlinear eigenfrequency spectrum leads to the loss-induced quantum revival of single-PB. This underlying principle is different from that of loss-induced entanglement~\cite{Plenio1999Cavity} in which a quantum effect is realized through conditional dynamics.

Figure~\ref{fig:FP3} shows that different types of quantum statistics can be tuned by increasing loss for the light revived after $\mathrm{CP}_{\mathrm{c}}$. As single-PB featuring two-photon antibunching, two-PB features three-photon antibunching, but with two-photon bunching, which indicates the absorption of two photons can suppress the absorption of additional photons~\cite{miranowicz2013Twophoton}. This two-PB effect can be characterized by the conditions $g^{(3)}_1(0)<1$ and $g^{(2)}_1(0)>1$, with $g^{(3)}_1(0)={\langle\hat{a}_{1}^{\dagger3}\hat{a}_{1}^{3}\rangle}/{\langle\hat{a}_{1}^{\dagger}\hat{a}_{1}\rangle^3}$~\cite{hamsen2017TwoPhoton}.

When the light recovers after $\mathrm{CP}_{\mathrm{c}}$, a two-PB emerges with $g^{(3)}_1(0)\sim 0.27$ and $g^{(2)}_1(0)\sim 1.12$ at $\gamma_{\mathrm{tip}}/\gamma_1^{'}=6$ [Fig.~\ref{fig:FP3}(a)]. Adding $\gamma_{\mathrm{tip}}$ beyond $\mathrm{CP}_{\mathrm{q}\uparrow}$ leads to a single-PB occurs at the EP. These results can also be confirmed by comparing the photon-number distribution $P_m$ with the Poisson distribution $\mathcal{P}_m$ [Fig.~\ref{fig:FP3}(b)]. We find that $P_2$ is enhanced while $P_{m>2}$ are suppressed at $\gamma_{\mathrm{tip}}/\gamma_1^{'}=6$, which is in sharp contrast to the case at the EP. With such a device, a switching between two-PB and single-PB can be achieved by increasing loss below or beyond $\mathrm{CP}_{\mathrm{q}\uparrow}$. As for as we know, this loss-induced quantum switching between different types of non-classical statistics has not been revealed in previous works on loss-induced classical revival~\cite{Guo2009Observation,Zhang2018Loss,peng2014Lossinduced}.

In summary, we have shown how to realize loss-induced quantum revival of single-PB in a compound nonlinear system. In contrast to the single-PB effects in conventional systems, we find less loss annihilates single-PB, and more loss helps to recover single-PB in quantum revival regime of light. This counterintuitive quantum effect happens because of the interplay of two-photon resonance and excitation-spectrum mode coalescence. More interestingly, different types of quantum correlations are exhibited in the revived light, which can be well controlled by tuning loss. These results, shedding light on the marriage of non-Hermitian physics and quantum optics at the single-photon levels, open up the way to reverse the effect of loss for steering quantum effects in various systems, such as plasmonics, metamaterials, and topological photonics. Our scheme no longer relies on destructive interference between different modes~\cite{Majumdar2012Loss,Huang2021Loss}, or additional gain media~\cite{peng2014Parity,Lin2016Loss}, which may enable novel quantum devices assisted by the loss for the applications of quantum engineering or metrology.

\begin{acknowledgements}
\emph{Acknowledgements.}---H.J. is supported by the National Natural Science Foundation of China (Grants No. 11935006 and No. 11774086). R.H. is supported by the Science and Technology Innovation Program of Hunan Province (Grant No. 2020RC4047). L.-M.K. is supported by the NSFC (Grants No. 11935006 and No. 11775075). X.-W.X. is supported by the NSFC (Grant No. 12064010) and the Natural Science Foundation of Hunan Province of China
 (Grant No. 2021JJ20036).
\end{acknowledgements}


%


\newpage

\onecolumngrid

\setcounter{equation}{0} \setcounter{figure}{0}
\setcounter{table}{0}
\setcounter{page}{1}\setcounter{secnumdepth}{3} \makeatletter
\renewcommand{\theequation}{S\arabic{equation}}
\renewcommand{\thefigure}{S\arabic{figure}}
\renewcommand{\bibnumfmt}[1]{[S#1]}
\renewcommand{\citenumfont}[1]{S#1}
\renewcommand\thesection{S\arabic{section}}

\begin{center}
{\large \bf Supplementary Material for ``Loss-Induced Quantum Revival''}
\end{center}

\begin{center}
Yunlan Zuo$^{1,^*}$, Ran Huang$^{1,2,^*}$, Le-Man Kuang$^{1,^\dagger}$, Xun-Wei Xu$^{1,^\ddagger}$, and Hui Jing$^{1,^\mathsection}$
\end{center}

\begin{minipage}[]{18cm}
\small{\it
\centering $^{1}$Key Laboratory of Low-Dimensional Quantum Structures and Quantum Control of Ministry of Education,  \\
\centering Department of Physics and Synergetic Innovation Center for Quantum  Effects and Applications, \\
\centering Hunan Normal University, Changsha 410081, China \\
\centering $^{2}$Theoretical Quantum Physics Laboratory, RIKEN Cluster for Pioneering Research, Wako-shi, Saitama 351-0198, Japan \\}

\end{minipage}

\vspace{8mm}
Here, we present more technical details on the intracavity field intensities and quantum correlation functions (Sec.~\ref{QCF}), as well as the cavity excitation spectrum and the eigensystem (Sec.~\ref{spectrum}).


\section{Intracavity field intensities and quantum correlation functions}\label{QCF}

We consider an optical-molecule system consisting of a Kerr resonator ($\mu\mathrm{R}1$) directly coupled to a linear resonator ($\mu\mathrm{R}2$). In a frame rotating with the driving frequency $\omega_{l}$, this system can be described by the following Hamiltonian
\begin{equation}
\hat{H}_{\mathrm{r}}=\Delta(\hat{a}_{1}^{\dagger}\hat{a}_{1}+\hat{a}_{2}^{\dagger}\hat{a}_{2})+\chi\hat{a}_{1}^{\dagger}\hat{a}_{1}^{\dagger}\hat{a}_{1}\hat{a}_{1}+J(\hat{a}_{1}^{\dagger}\hat{a}_{2}+\hat{a}_{2}^{\dagger}\hat{a}_{1})+\Omega(\hat{a}_{1}^{\dagger}+\hat{a}_{1}),
\end{equation}
where $\Delta=\omega_{c}-\omega_{l}$ is the optical detuning, $\hat{a}_{j=1,2}$ are the intracavity modes with resonance frequency $\omega_{c}$, $J$ is the coupling strength between the two resonators, $\chi=3\hbar\omega_{c}^{2}\chi^{(3)}/(4\varepsilon_{0}\varepsilon_{r}^{2}V_{\mathrm{eff}})$ is the Kerr parameter with vacuum (relative) permittivity $\varepsilon_{0}$ ($\varepsilon_{r}$), nonlinear susceptibility $\chi^{(3)}$, and mode volume $V_{\mathrm{eff}}$. The driving amplitude is given by $\Omega=\sqrt{\gamma_{\mathrm{ex}}P_{\mathrm{in}}/(\hbar\omega_{l})}$ with the power $P_{\mathrm{in}}$ on $\mu\mathrm{R}1$, and the loss induced by the coupling between the resonator and the fiber taper $\gamma_{\mathrm{ex}}$.

The optical decay can be included in the effective Hamiltonian $\hat{H}_{\mathrm{eff}}=\hat{H}_{\mathrm{r}}-i{\sum}_{j=1,2}(\gamma_{j}^{'}/2)\hat{a}_{j}^{\dagger}\hat{a}_{j}$,
where $\gamma_{1}^{'}=\gamma_{1}+\gamma_{\mathrm{ex}}$ ($\gamma_{2}^{'}=\gamma_{2}+\gamma_{\mathrm{tip}}$) is the total loss of $\mu\mathrm{R}1$ ($\mu\mathrm{R}2$), $\gamma_{1}$ and $\gamma_{2}$ are the intrinsic losses of the two resonators, and an additional loss $\gamma_{\mathrm{tip}}$ is induced on $\mu\mathrm{R}2$ by a chromium (Cr) coated silica-nanofiber tip. Under the weak-driving condition ($\Omega\ll\gamma_{1}^{'}$), the Hilbert space can be restricted to a subspace with few photons. In the subspace with $N=m+n=3$ excitations, the general state of the system can be expressed as
\begin{equation}\label{SM_state}
|\psi(t)\rangle =\sum\limits_{N=0}^{3}\sum\limits_{m=0}^{N}C_{m,N-m}|m,N-m\rangle,
\end{equation}
with probability amplitudes $C_{m,N-m}$, which can be obtained by solving the Schr\"{o}dinger equation:
\begin{equation}
i|\dot{\psi}(t)\rangle =\hat{H}_{\mathrm{eff}}|\psi(t)\rangle.
\end{equation}

\begin{figure*}[t]
\centering
\includegraphics[width=0.9 \textwidth]{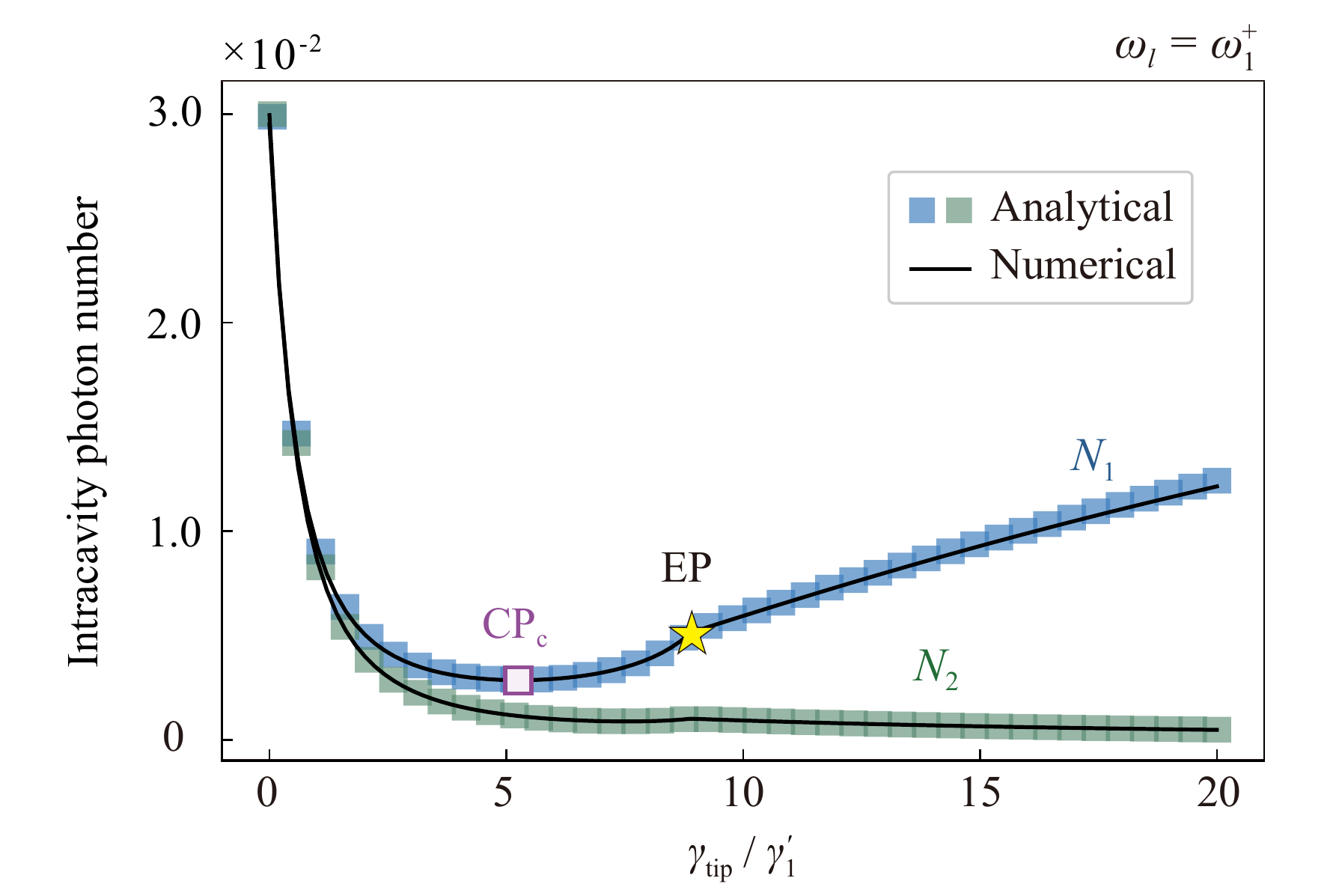}
\caption{Intracavity photon numbers $N_{1}$ (blue) and $N_{2}$ (green) versus $\gamma_{\mathrm{tip}}$. Below $\mathrm{CP_{c}}$, $N_{1}$ and $N_{2}$ decrease by increasing $\gamma_{\mathrm{tip}}$. When $\gamma_{\mathrm{tip}}$ exceeds $\mathrm{CP_{c}}$, $N_{1}$ is revived, while $N_{2}$ keep decreasing, resulting in a predominant mode localized in $\mu\mathrm{R}1$. The analytical results (colored squares) agree well with the numerical results (black solid curves). The parameter are the same as those in the main text.} \label{FS1}
\end{figure*}

\begin{figure*}[t]
\centering
\includegraphics[width=0.96 \textwidth]{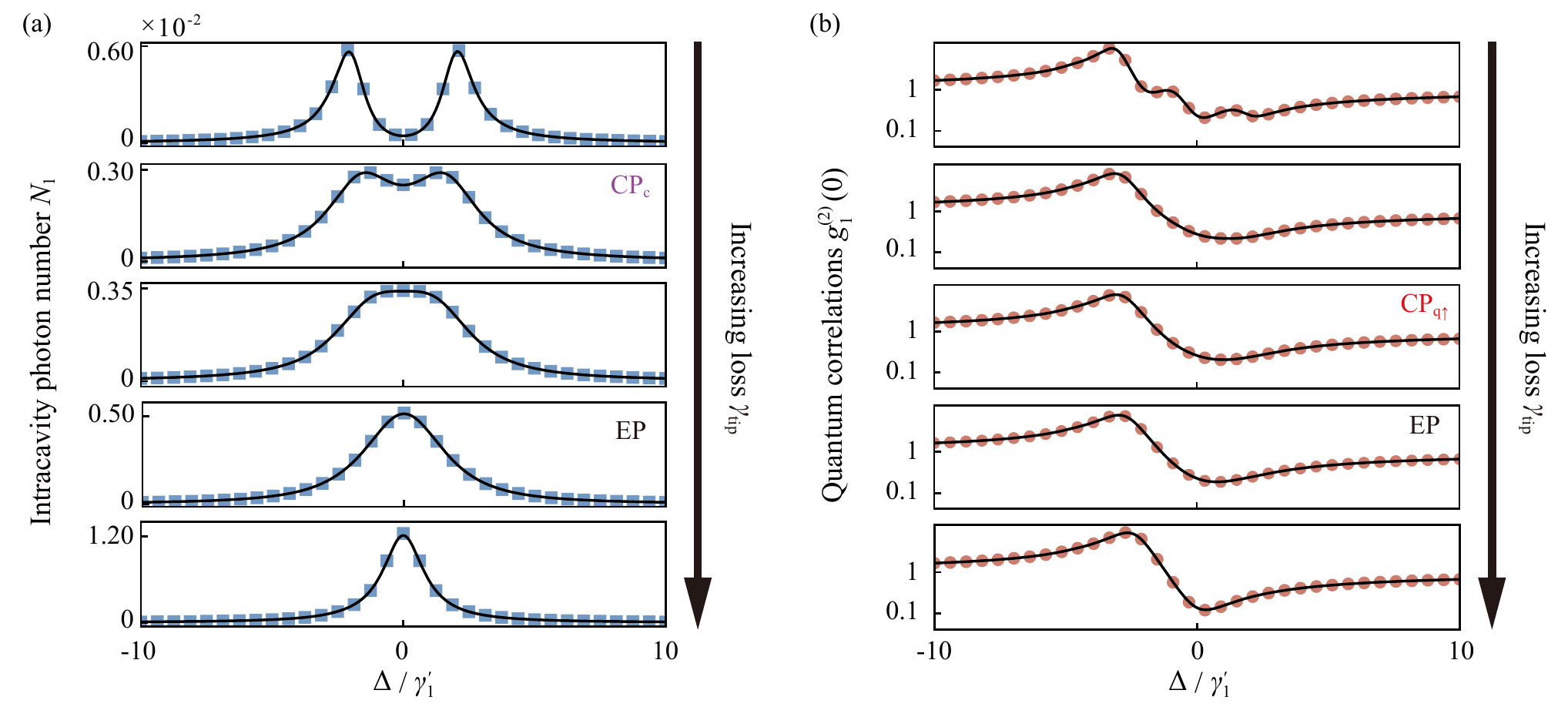}
\caption{(a) The intracavity photon number $N_{1}$ and (b) second-order correlation $g_{1}^{(2)}(0)$ versus the optical detuning $\Delta$ for different $\gamma_{\mathrm{tip}}$. The analytical results (colored markers) agree well with the numerical results (black solid curves). The parameters are the same as those in Fig.~\ref{FS1}. } \label{FS2}
\end{figure*}

When a weak-driving field is applied to the cavity, it may excites few photons in the cavity. Thus, we can approximate the probability amplitudes of the excitations as $C_{m,N-m}\sim(\Omega/\gamma_{1}^{'})^{N}$. By using a perturbation method and discarding higher-order terms in each equation for lower-order variables, we obtain the following equations of motion for the probability amplitudes
\begin{align}
i\dot{C}_{00}(t) & =0, \qquad i\dot{C}_{01}(t)=\Delta_{2}C_{01}(t)+JC_{10}(t), \qquad \;\: i\dot{C}_{10}(t)=\Delta_{1}C_{10}(t)+JC_{01}(t)+\Omega C_{00}(t),\nonumber \\
i\dot{C}_{02}(t) & =2\Delta_{2}C_{02}(t)+\sqrt{2}JC_{11}(t), \qquad \qquad \qquad \qquad i\dot{C}_{20}(t)=2\Delta_{3}C_{20}(t)+\sqrt{2}JC_{11}(t)+\sqrt{2}\Omega C_{10}(t),\nonumber \\
i\dot{C}_{11}(t) & =(\Delta_{1}+\Delta_{2})C_{11}(t)+\sqrt{2}JC_{20}(t)+\sqrt{2}JC_{02}(t)+\Omega C_{01}(t),\nonumber \\
i\dot{C}_{03}(t) & =3\Delta_{2}C_{03}(t)+\sqrt{3}JC_{12}(t), \qquad \qquad \qquad \qquad i\dot{C}_{12}(t)=\Delta_{6}C_{12}(t)+2JC_{21}(t)+\sqrt{3}JC_{03}(t)+\Omega C_{02}(t),\nonumber \\
i\dot{C}_{30}(t) & =3\Delta_{4}C_{30}(t)+\sqrt{3}JC_{21}(t)+\sqrt{3}\Omega C_{20}(t), \qquad i\dot{C}_{21}(t)=\Delta_{5}C_{21}(t)+\sqrt{3}JC_{30}(t)+2JC_{12}(t)+\sqrt{2}\Omega C_{11}(t),\nonumber \\
\end{align}
where $\Delta_{1}=\Delta-i\gamma_{1}^{'}/2$, $\Delta_{2}=\Delta-i\gamma_{2}^{'}/2$,
$\Delta_{3}=\Delta_{1}+\chi$, $\Delta_{4}=\Delta_{1}+2\chi$, $\Delta_{5}=2\Delta_{3}+\Delta_{2}$ and $\Delta_{6}=\Delta_{1}+2\Delta_{2}$. For the initially empty resonators, i.e., the initial state of the
system is the vacuum state $\left|00\right\rangle $, the initial
condition reads as $C_{00}(0)=1$. By setting $\dot{C}_{mn}(t)=0$, we obtain the following solutions
\begin{align}
C_{01} & =\frac{J\Omega}{\eta_{1}}, \qquad \quad C_{10}=-\frac{\Omega\Delta_{2}}{\eta_{1}}, \qquad \;\: C_{02}=\frac{\sqrt{2}\Omega^{2}J^{2}(\Delta_{3}+\Delta_{2})}{\eta_{1}\eta_{2}}, \nonumber \\
C_{20} & =\frac{\sqrt{2}\Omega^{2}\Delta_{2}^{2}(\Delta_{1}+\Delta_{2})}{\eta_{1}\eta_{2}}, \qquad \qquad \quad C_{11}=\frac{-2\Omega^{2}\Delta_{2}J(\Delta_{3}+\Delta_{2})}{\eta_{1}\eta_{2}}, \nonumber \\
C_{03} & =\frac{-\sqrt{6}J^{3}\Omega^{3}[\xi_{2}(\Delta_{2}+\Delta_{3})-2\Delta_{2}^{2}(\Delta_{1}+\Delta_{2})]}{3\eta_{1}\eta_{2}\mu},\nonumber \\
C_{12} & =\frac{\sqrt{2}J^{2}\Omega^{3}\Delta_{2}[\xi_{2}(\Delta_{2}+\Delta_{3})-2\Delta_{2}^{2}(\Delta_{1}+\Delta_{2})]}{\eta_{1}\eta_{2}\mu},\nonumber \\
C_{30} & =\frac{\sqrt{6}\Omega^{3}[\Delta_{2}^{2}(4J^{2}\Delta_{2}+\Delta_{5}\eta_{3})(\Delta_{1}+\Delta_{2})-2J^{2}\Delta_{2}^{2}\Delta_{6}(\Delta_{2}+\Delta_{3})]}{3\eta_{1}\eta_{2}\mu},\nonumber \\
C_{21} & =-\frac{\sqrt{2}J\Omega^{3}[\Delta_{2}^{2}\eta_{3}(\Delta_{1}+\Delta_{2})-2\Delta_{2}^{2}\Delta_{4}\Delta_{6}(\Delta_{2}+\Delta_{3})]}{\eta_{1}\eta_{2}\mu},
\end{align}
where $\eta_{1}=\Delta_{1}\Delta_{2}-J^{2}$, $\eta_{2}=2\xi_{1}\Delta_{2}-2J^{2}\Delta_{3}$, $\eta_{3}=J^{2}-\Delta_{2}\Delta_{6}$, $\xi_{1}=\Delta_{1}\Delta_{3}+\Delta_{2}\Delta_{3}-J^{2}$, $\xi_{2}=J^{2}-4\Delta_{2}\Delta_{4}-\Delta_{4}\Delta_{5}$ and $\mu=J^{2}\xi_{2}-J^{2}\Delta_{2}\Delta_{6}+\Delta_{2}\Delta_{4}\Delta_{5}\Delta_{6}$. The probabilities of finding $m$ photons in $\mu\mathrm{R}1$ and
$n$ photons in $\mu\mathrm{R}2$ are given by $P_{mn}=|C_{mn}|^{2}$. The mean-photon numbers in $\mu\mathrm{R}1$ and $\mu\mathrm{R}2$ are denoted by $N_{1}$ and $N_{2}$, respectively, and can be obtained from the above probability distribution as
\begin{align}\label{SM_N}
N_{1} & =\langle \hat{a}_{1}^{\dagger}\hat{a}_{1}\rangle=\sum\limits_{N=0}^{3}\sum\limits_{m=0}^{N}mP_{mn}, \qquad N_{2}=\langle \hat{a}_{2}^{\dagger}\hat{a}_{2}\rangle=\sum\limits_{N=0}^{3}\sum\limits_{n=0}^{N}nP_{mn}.
\end{align}
The equal-time (namely zero-time-delay) second-order correlation function of $\mu\mathrm{R}1$ is written as
\begin{equation}
g_{1}^{(2)}(0)=\frac{\langle \hat{a}_{1}^{\dagger2}\hat{a}_{1}^{2}\rangle }{\langle \hat{a}_{1}^{\dagger}\hat{a}_{1}\rangle ^{2}}=\frac{\langle \hat{m}^{2}-\hat{m}\rangle}{\langle \hat{m}\rangle ^{2}}=\frac{2P_{20}+6P_{30}+2P_{21}}{N_{1}^{2}}\simeq\frac{4\eta_{1}^{2}(\Delta_{1}+\Delta_{2})^{2}}{\eta_{2}^{2}}.
\end{equation}
The approximate equal-time third-order correlation function is written as
\begin{equation}
g_{1}^{(3)}(0)=\frac{\langle \hat{a}_{1}^{\dagger3}\hat{a}_{1}^{3}\rangle }{\langle \hat{a}_{1}^{\dagger}\hat{a}_{1}\rangle ^{3}}=\frac{\langle \hat{m}^{3}-3\hat{m}^{2}+2\hat{m}\rangle }{\langle \hat{m}\rangle ^{3}}=\frac{6P_{30}}{N_{1}^{3}}\simeq\frac{4\eta_{1}^{4}\{ \Delta_{2}^{2}(4J^{2}\Delta_{2}+\Delta_{5}\eta_{3})(\Delta_{1}+\Delta_{2})-2J^{2}\Delta_{2}^{2}\Delta_{6}(\Delta_{2}+\Delta_{3})\} ^{2}}{\eta_{2}^{2}\mu^{2}\Delta_{2}^{6}}.
\end{equation}
An excellent agreement between our analytical results and the exact numerical results is shown in Figs.~\ref{FS1} and \ref{FS2}.


\section{Cavity excitation spectrum and eigensystem}\label{spectrum}

\begin{figure*}[t]
\centering
\includegraphics[width=0.96 \textwidth]{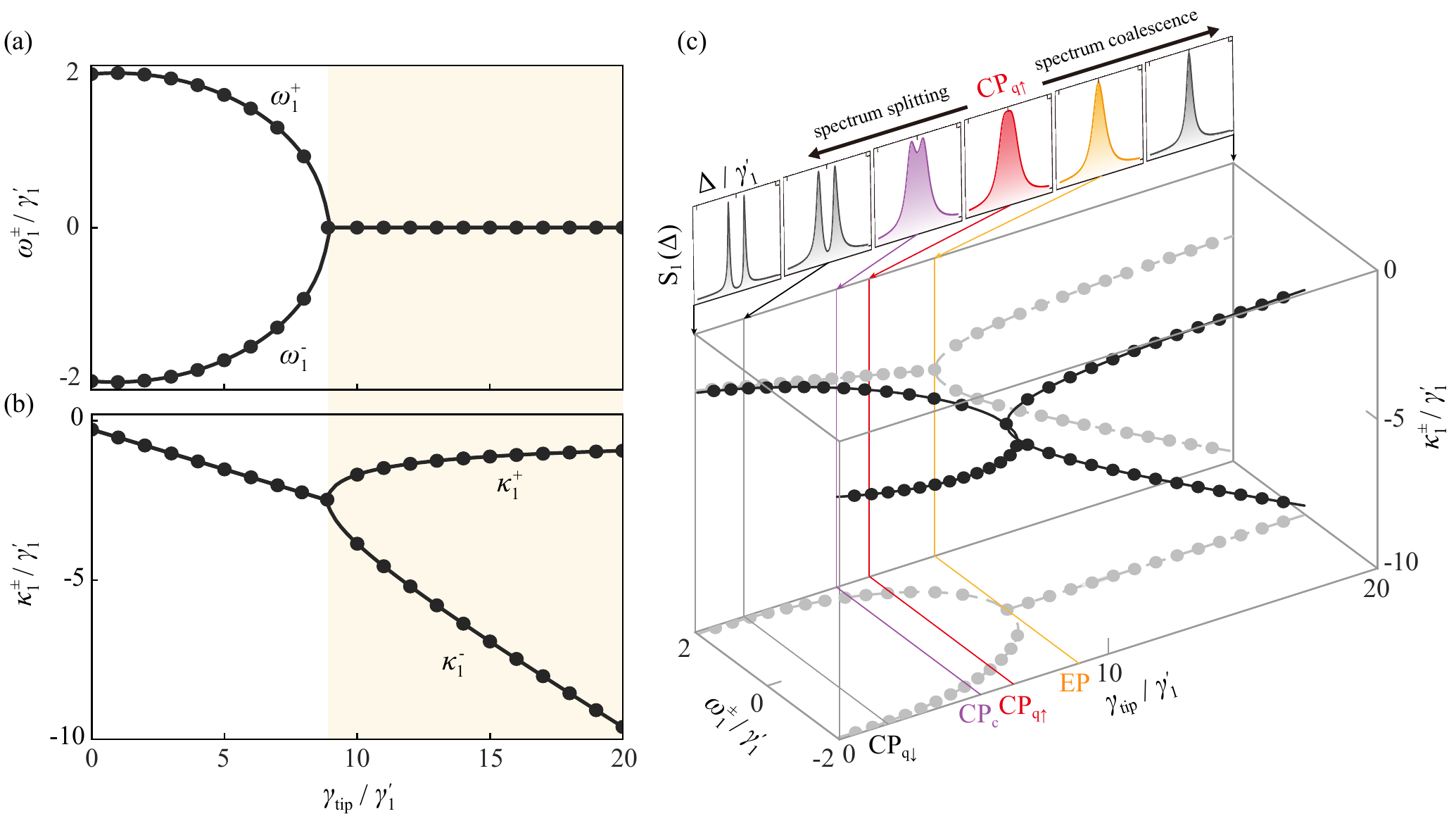}
\caption{(a) The real parts $\omega_{1}^{\pm}$ and (b) imaginary parts $\kappa_{1}^{\pm}$ of the eigenvalues $\lambda_{1}^\pm$ as a function of $\gamma_{\mathrm{tip}}$. (c) The evolutions of the excitation spectrum $S_{1} (\Delta)$ and eigenvalues versus $\gamma_{\mathrm{tip}}$. The $S_{1} (\Delta)$ shows the spectrum splitting and spectrum coalescence. The parameters are the same as those in Fig.~\ref{FS1}} \label{FS3}
\end{figure*}

The loss-induced quantum and switch require the interplay of the mode coalescence in cavity excitation spectrum and the two-photon resonance in nonlinear eigenenergy structure. The excitation spectrum of the $\mu\mathrm{R}1$ is given by
\begin{equation}
S_{1}(\Delta)=\frac{N_{1}}{n_{0}},
\end{equation}
where $n_{0}=\Omega^{2}/(\gamma_{1}^{'}+\gamma_{2}^{'})^{2}$ is the normalization factor. The nonlinear eigenenergy spectrum can be obtained through the following Hamiltonian:
\begin{equation}\label{SM_He}
\hat{H}_{\mathrm{e}}=\hat{H}_{\mathrm{i}}-i\frac{\gamma_{1}^{'}}{2}\hat{a}_{1}^{\dagger}\hat{a}_{1}-i\frac{\gamma_{2}^{'}}{2}\hat{a}_{2}^{\dagger}\hat{a}_{2},
\end{equation}
where $\hat{H}_{\mathrm{i}}=\omega_{c}(\hat{a}_{1}^{\dagger}\hat{a}_{1}+\hat{a}_{2}^{\dagger}\hat{a}_{2})+\chi\hat{a}_{1}^{\dagger}\hat{a}_{1}^{\dagger}\hat{a}_{1}\hat{a}_{1}+J(\hat{a}_{1}^{\dagger}\hat{a}_{2}+\hat{a}_{2}^{\dagger}\hat{a}_{1})$ is the Hamiltonian of the isolated system. Since $[\hat{a}_{1}^{\dagger}\hat{a}_{1}+\hat{a}_{2}^{\dagger}\hat{a}_{2},\hat{H}_{\mathrm{e}}]=0$, i.e., the total excitation number is conserved, we can obtain the eigensystem with the Hilbert space spanned by the basis state $|m,n\rangle $, i.e., the Fock state with $m$ photons in $\mu\mathrm{R}1$ and $n$ photons in $\mu\mathrm{R}2$.

\begin{figure*}[t]
\centering
\includegraphics[width=0.96 \textwidth]{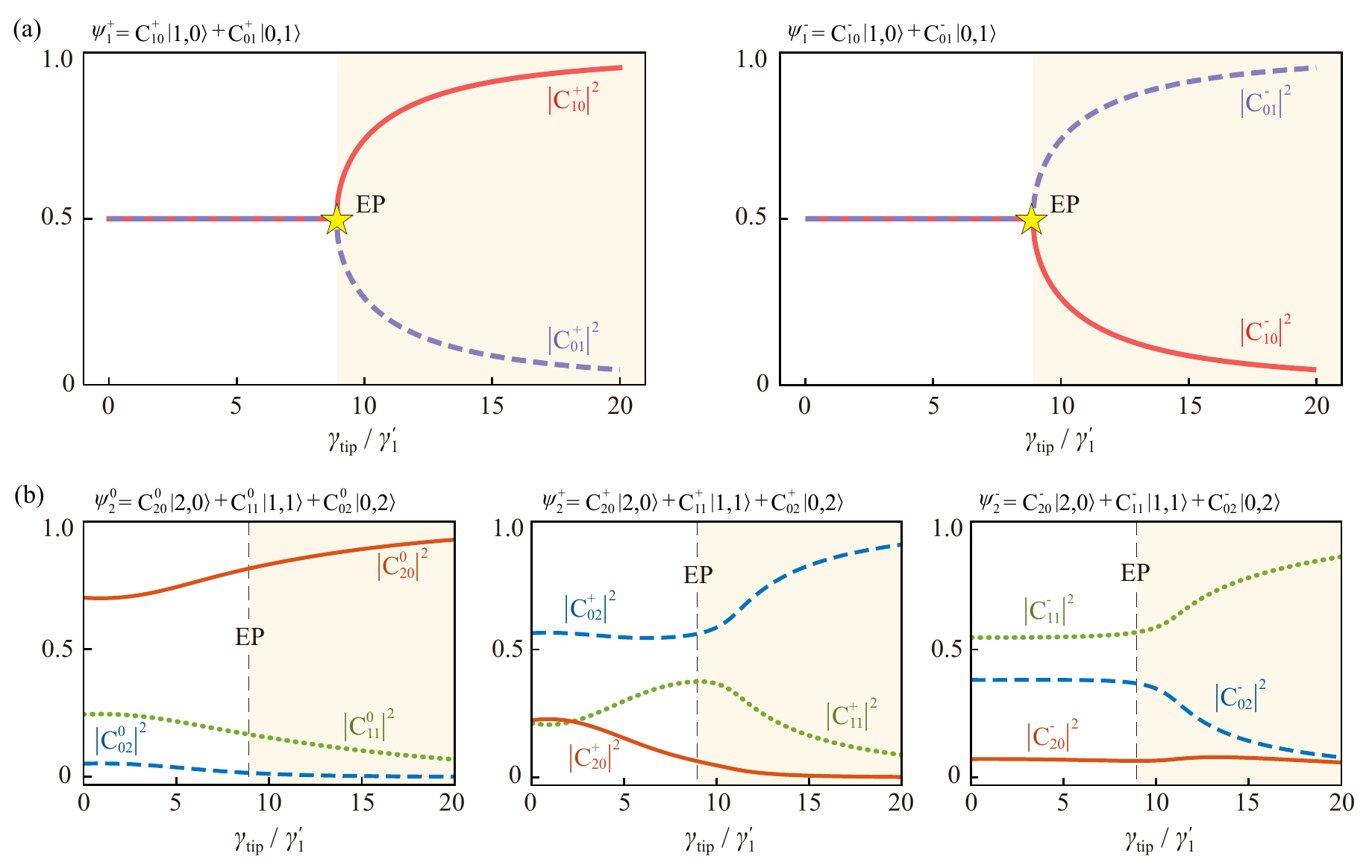}
\caption{After the EP, (a) the single-photon eigenstates $|\psi_{1} ^{\pm}\rangle$ are intensively localized on $|1,0\rangle$ and $|0,1\rangle$, respectively; (b) The $|\psi_{2} ^{0,\pm}\rangle$ are respectively governed by the states $|2,0\rangle$, $|0,2\rangle$ and $|1,1\rangle$. The parameters are the same as those in Fig.~\ref{FS1}} \label{FS4}
\end{figure*}

For the subspace with zero photons, we have $\hat{H}_{\mathrm{e}}\psi_{0}=\lambda_{0}\psi_{0}$, and the eigenstate is given by $\psi_{0}=\left|0,0\right\rangle $ with the eigenvalue $\lambda_{0}=0$. In this subspace with one photon, the Hamiltonian can be expressed as
\begin{equation}
\hat{H}_{\mathrm{e}}=\left(\begin{array}{cc}
\omega_{c}-i\frac{\gamma_{1}^{'}}{2} & J\\
J & \omega_{c}-i\frac{\gamma_{2}^{'}}{2}
\end{array}\right).
\end{equation}
The eigenvalues are $\lambda_{1}^{\pm}=-i\Gamma+\omega_{c}\pm\sqrt{J^{2}-\beta^{2}}$, whose real and imaginary parts are respectively indicate the eigenfrequencies $\omega_{1}^{\pm}$ and the linewidths $\kappa_{1}^{\pm}$. Here, $\Gamma=(\gamma_{1}^{'}+\gamma_{2}^{'})/4$ and $\beta=(\gamma_{2}^{'}-\gamma_{1}^{'})/4$ quantify the total loss and the loss contrast of the system, respectively. The Hamiltonian exceptional points (HEPs) are emerge for $\lambda_{1}^{+}=\lambda_{1}^{-}$, i.e., $\gamma_{\mathrm{tip}}^{\mathrm{EP}}=4J+\gamma_{1}^{'}-\gamma_{2}$. The corresponding eigenstates are $\psi_{1}^{\pm}=C_{10}^{\pm}|1,0\rangle+C_{01}^{\pm}|0,1\rangle$,
where $C_{10}^{\pm}=J\mathcal{N}_{1}^{\pm}$, $C_{01}^{\pm}=-(i\beta\mp\sqrt{J^{2}-\beta^{2}})\mathcal{N}_{1}^{\pm}$, and $\mathcal{N}_{1}^{\pm}=(|J|^{2}+|i\beta\mp\sqrt{J^{2}-\beta^{2}}|^{2})^{-1/2}$.

In this subspace with two photons, we express the Hamiltonian in the matrix form as
\begin{equation}
\hat{H}_{\mathrm{e}}=\left(\begin{array}{ccc}
2\omega_{c}+2\chi-i\gamma_{1}^{'} & \sqrt{2}J & 0\\
\sqrt{2}J & 2\omega_{c}-i\frac{\gamma_{1}^{'}+\gamma_{2}^{'}}{2} & \sqrt{2}J\\
0 & \sqrt{2}J & 2\omega_{c}-i\gamma_{2}^{'}
\end{array}\right).
\end{equation}
By solving the characteristic equation, we find the eigenvalues as
\begin{align}
\lambda_{2}^{0} =G-\frac{(1-i\sqrt{3})E}{3\times2^{2/3}F}+\frac{(1+i\sqrt{3})F}{6\times2^{1/3}},\ \ 
\lambda_{2}^{+} =G-\frac{(1+i\sqrt{3})E}{3\times2^{2/3}F}+\frac{(1-i\sqrt{3})F}{6\times2^{1/3}}, \ \ 
\lambda_{2}^{-} =G+\frac{2^{1/3}E}{3F}-\frac{F}{3\times2^{1/3}},
\end{align}
where 
\begin{align}
A&=2\omega_{c}+2\chi-i\gamma_{1}^{'},\quad B=2\omega_{c}-i{\gamma_{1}^{'}+\gamma_{2}^{'}}/{2}, \quad C=2\omega_{c}-i\gamma_{2}^{'},\nonumber \\
D&=36J^{2}\chi+{9}\chi(\gamma_{1}^{'}-\gamma_{2}^{'})^{2}/2-16\chi^{3}+i18\chi^{2}(\gamma_{1}^{'}-\gamma_{2}^{'}), \nonumber \\
E&=-12J^{2}+{3}(\gamma_{1}^{'}-\gamma_{2}^{'})^{2}/4-4\chi^{2}+i3\chi(\gamma_{1}^{'}-\gamma_{2}^{'}),\nonumber \\
F&=[D+\sqrt{4E^{3}+D^{2}}]^{1/3},\quad G={1}(A+B+C)/3.
\end{align}
The corresponding eigenstates are $\psi_{2}^{\pm,0}=C_{20}^{\pm,0}|2,0\rangle+C_{11}^{\pm,0}|1,1\rangle +C_{02}^{\pm,0}|0,2\rangle$, where $C_{20}^{\pm,0}=\sqrt{2}J(C-\lambda_{2}^{\pm,0})\mathcal{N}_{2}^{\pm,0}$, $C_{11}^{\pm,0}=-(C-\lambda_{2}^{\pm,0})|A-\lambda_{2}^{\pm,0}|\mathcal{N}_{2}^{\pm,0}$, and $C_{02}^{\pm,0}=\sqrt{2}J|A-\lambda_{2}^{\pm,0}|\mathcal{N}_{2}^{\pm,0}$.

Hamiltonian EPs do not take into account the quantum noise associated with quantum jumps. For a full quantum picture, one should resort to the EPs of the system's Liouvillian. This can be done using the Lindblad master-equation approach, and the Liouvillian superoperator $\mathscr{\mathcal{L}}$ is given by~\cite{Minganti2019quantum}
\begin{equation}
\mathcal{L}\hat{\rho}=-i[\hat{H}_{\mathrm{i}},\hat{\rho}]+\underset{j=1,2}{\sum}\mathcal{D}(\hat{\rho},\hat{A}_{j}),
\end{equation}
where $\mathcal{D}(\hat{\rho},\hat{A}_{j})=\hat{A}_{j}\hat{\rho}\hat{A}_{j}^{\dagger}-\hat{A}_{j}^{\dagger}\hat{A}_{j}\hat{\rho}/2-\hat{\rho}\hat{A}_{j}^{\dagger}\hat{A}_{j}/2$ are the dissipators associated with the jump operators $\hat{A}_{j}=\sqrt{\gamma_{j}^{'}}\hat{a}_{j}$. We then find the Liouvillian EPs (LEPs) as the degeneracies of the Liouvillian superoperator by solving the equation~\cite{Minganti2019quantum}: $\mathscr{\mathcal{L}}\hat{\rho}_{i}=\varLambda_{i}\hat{\rho}_{i}$, where $\varLambda_{i}$ and $\hat{\rho}_{i}$ are the eigenvalues and the corresponding eigenstates of $\mathscr{\mathcal{L}}$. As a result, the LEPs and HEPs occur at the same positions indicating a good agreement between the semiclassical and fully quantum approaches \cite{Minganti2019quantum}.

The cavity excitation spectrum becomes coalescent at the quantum critical point $\mathrm{CP}_{\mathrm{q}\uparrow}$ (Fig.~\ref{FS3}). The eigenstates $|\psi_{1} ^{\pm}\rangle$ are respectively governed by the states $|1,0\rangle$ and $|0,1\rangle$ when the system beyond the EP [Fig.~\ref{FS4}(a)]. Figure \ref{FS4}(b) shows that the $|\psi_{2} ^{\pm,0}\rangle$ are respectively governed by the states $|0,2\rangle$, $|1,1\rangle$ and $|2,0\rangle$ when the system operates at or beyond the EP.






%

\end{document}